\documentclass[journal]{IEEEtran}
\usepackage[utf8]{inputenc}

\usepackage{xcolor}
\usepackage{color, soul}
\usepackage{enumerate}
\usepackage[shortlabels]{enumitem}
\usepackage{hyperref}
\usepackage{cite}
\usepackage{tabu}
\usepackage{bm, bbm} 
\usepackage{mathtools}
\usepackage{amsmath, amssymb, amsthm}
\usepackage{nicefrac}
\usepackage{empheq}

\usepackage{booktabs}
\usepackage{graphicx}
\usepackage{multirow}
\usepackage{units}
\usepackage{siunitx}
\usepackage{enumitem}
\usepackage{hyperref}
\usepackage{svg}

\usepackage{amsmath,amsfonts,amsthm}

\usepackage[noabbrev,capitalize]{cleveref}

\crefname{equation}{}{}
\Crefname{equation}{}{}

\theoremstyle{definition} 
\newtheorem{prop}{Proposition}

\theoremstyle{plain} 

\theoremstyle{remark} 
\newtheorem{remark}{Remark}

\usepackage{tikz}
\newcommand*\circled[1]{\tikz[baseline=(char.base)]{
            \node[shape=circle,draw,thin,inner sep=0.5pt] (char) {#1};}}

\newcommand{\set}[1]{\mathcal{#1}} 
\newcommand{\norm}[1]{\left\lVert#1\right\rVert} 

\newcommand{\om}{\bm{\omega}}

\DeclareMathOperator{\Var}{Var}

\DeclareMathOperator{\Eptn}{\mathbb{E}}
\DeclareMathOperator{\RM}{\mathbb{F}}
\DeclareMathOperator{\Prb}{\mathbb{P}}

\DeclareMathOperator{\Sigmart}{\Sigma^{\nicefrac{1}{2}}}
\DeclareMathOperator{\Sigmartk}{\Sigma_k^{\nicefrac{1}{2}}}
\DeclareMathOperator{\conv}{conv}

\def\myproofstart{\noindent\textit{Proof. }}
\def\myproofend{
    \hspace*{\fill} $\square$
}

\usepackage{placeins}
\newcommand{\subparagraph}{}
\usepackage{titlesec}

\titlespacing*{\section}{0pt}{5pt}{3pt}
\titlespacing*{\subsection}{0pt}{3pt}{0pt}
\titlespacing*{\subsubsection}{0pt}{3pt}{0pt}
\title{Risk Trading in a Chance-Constrained Stochastic Electricity Market}
\author{Robert Mieth, Matt Roveto, and Yury Dvorkin. 
\vspace*{-7mm}
}

\begin{document}

\pagestyle{empty}
\bstctlcite{IEEE:BSTcontrol} 

\maketitle

\begin{abstract}
Existing electricity market designs assume risk neutrality and lack risk-hedging instruments, which leads to suboptimal market outcomes and reduces the overall market efficiency. 
This paper enables  risk-trading in the chance-constrained stochastic electricity market by introducing Arrow-Debreu Securities (ADS) and derives a risk-averse market-clearing model with risk trading. To  enable risk trading, the probability space of underlying uncertainty is discretized in a finite number of outcomes,  which makes it possible to design practical risk contracts and  to produce energy, balancing reserve and risk prices. Notably, although risk contracts are discrete, the model preserves the continuity of chance constraints. The case study illustrates the usefulness of the proposed risk-averse chance-constrained electricity market with risk trading.
\vspace{-2mm}
\end{abstract}

\section{Introduction}

Uncertain renewable energy sources (RES) challenge the efficiency of existing wholesale electricity markets, which still lack risk-hedging financial instruments, \cite{bose2019some}. 
As a result, electricity markets are incomplete with respect to uncertainty and risk, i.e. they do not provide market participants with a mechanism to secure their positions relative to all probable future states of the system. 
{\color{black} 
Motivated by the previously developed chance-constrained optimal power flow formulation in \cite{bienstock2014chance}, we developed a \textit{chance-constrained} stochastic electricity market design in \cite{kuang2018pricing, dvorkin2019chancemarket,mieth2019risk,mieth2019distribution},
}%
which internalizes the RES uncertainty and  produces uncertainty-aware electricity prices that support welfare efficiency, revenue adequacy and cost recovery. 
{\color{black} 
However, \cite{kuang2018pricing, dvorkin2019chancemarket,mieth2019risk,mieth2019distribution} assume (i) risk-neutrality and (ii) a single common belief on the system uncertainty. 
In reality, market participants are likely to trade (i) in a risk-averse manner and (ii) with different uncertainty beliefs.
Thus, decisions are more conservative and lead to less efficient market outcomes if there is no opportunity to compensate the risk of uncertain cost with financial securities, \cite{ralph2011pricing}.}

Although common in the fields of stochastic optimization and finance \cite{rockafellar2007coherent}, the notion of risk aversion has only recently gained attention in power system operations and electricity markets. 
For example, Hans \textit{et al.} \cite{hans2019risk} developed risk-averse control strategies for decentralized generation resources and Kazempour \textit{et al.} \cite{kazempour2016effects} explored the effects of risk-averse electricity producers in a two-stage market equilibrium. 
However, while hedging uncertain cost against risk using the conditional value-at-risk (CVaR),  \cite{hans2019risk,kazempour2016effects} do not consider risk trading.  
{\color{black} 
On the other hand, building on the theoretical groundwork of Ralph and Smeers \cite{ralph2011pricing,ralph2015risk}, Philpott \textit{et al.} \cite{philpott2016equilibrium} proposed a risk-complete multi-stage scenario-based stochastic energy market by introducing risk-trading via Arrow-Debreu Securities (ADS). This risk completeness, i.e.  risk trading via financial instruments parallel to all other traded assets and services, provably enabled the existence of a risk-averse competitive equilibrium, if all market participants are endowed with a coherent risk measure. 
}
{Gérard \textit{et al.} \cite{gerard2018risk} applied the result from \cite{philpott2016equilibrium} to a two-stage stochastic electricity market and showed that a risk-averse equilibrium might not be unique.  
In line with \cite{philpott2016equilibrium,gerard2018risk}, Cory-Wright and Zakeri \cite{cory2018stochastic} demonstrated that different risk perceptions of market participants may encourage them to act strategically, thus causing suboptimal market outcomes, which can be avoided in risk-complete electricity markets. }

Departing from  scenario-based stochastic programming used in market designs in \cite{kazempour2018stochastic,martin2014stochastic,wong2007pricing,philpott2016equilibrium,kazempour2016effects,gerard2018risk}, this paper explores risk trading via ADS in the chance-constrained electricity market proposed in \cite{kuang2018pricing,dvorkin2019chancemarket,mieth2019risk,mieth2019distribution}. 
{\color{black} 
Unlike  data-demanding scenario-based approaches, chance constraints only require statistical moments to  internalize uncertainty in the market design using  continuous probability distributions.
Therefore, we first develop a general risk-complete chance constrained electricity market with continuous, infinite-dimensional ADS. Second, we show that ADS can be  discretized to enable  practical risk contracts for a given set of uncertain outcomes.
}
Finally, this paper   analyzes risk-averse market outcomes and investigates the effects of risk trading on market prices.  


\section{Chance-constrained  Electricity Market}

Consider a chance-constrained electricity market as in \cite{dvorkin2019chancemarket,mieth2019risk,kuang2018pricing}. Let $\set{N}$, $\set{G}$, $\set{U}$ be sets of nodes, conventional generators, and RES. The market operator solves:
\allowdisplaybreaks
\begin{subequations}
\begin{align}
     \min_{p_{G,i}, \alpha_i}  \quad & 
     \RM_0 \left[\sum_{i\in\set{G}} c_i(p_{G,i}(\om))\right] \label{base:objective}\\
    \text{s.t.} \quad 
    & p_{U,i}(\om_i) = p_{U,i} + \om_i && \forall i \in \set{U} \label{base:uncert_res}\\
    & p_{G,i}(\om) = p_{G,i} - \alpha_i^{\top} \om && \forall i \in \set{G} \label{base:uncert_generation}\\
    (\delta_i^+): \quad 
        & \Prb[p_{G,i}(\om) \leq \overline{p}_{G,i}] \geq 1-\epsilon_g && \forall i \in \set{G}  \label{base:upper_gen_cc}\\
    (\delta_i^-): \quad
        & \Prb[p_{G,i}(\om) \geq \underline{p}_{G,i}] \geq 1-\epsilon_g && \forall i \in \set{G} \label{base:lower_gen_cc}  \\
    (\theta): \quad
        & \Prb[F(p_{G}(\om), p_{U}(\om), p_{D}) \in \set{F}] \geq 1-\epsilon_f \hspace{-2cm}&& \label{base:flowconstraints} \\
    (\lambda_i): \quad
        &  p_{G,i} + p_{U,i} + p_i(F)  = p_{D,i} && \forall i \in \set{N}\label{base:enerbal}\\
    (\chi_u): \quad
        & \sum_{i\in\set{G}} \alpha_{i,u} = 1 && \forall u \in \set{U} \label{base:balbal},
\end{align}%
\label{mod:base-cc}%
\end{subequations}%
\allowdisplaybreaks[0]%
where  \cref{base:objective} minimizes the system operating cost evaluated by measure $\RM_0$ (e.g. expectation, if $\RM_0 \equiv \Eptn$) over the random vector of RES forecast errors $\om=[\om_i, i\in\set{U}]$ and given the cost function of each generator $c_i(p_{G,i})$. Eq.~\cref{base:uncert_res} models the uncertain RES power output $p_{U,i}(\om_i)$ at node $i$ as the RES forecast $p_{U,i}$ plus the  RES forecast error $\om_i$. Eq.~\cref{base:uncert_generation} defines the power output of conventional generators under uncertainty $p_{G,i}(\om)$ using an affine control policy, where  $p_{G,i}$  and  $\alpha_i = [0\leq  \alpha_{i,u}\leq 1, u \in \set{U}]$ are decisions for the scheduled power output and the vector of participation factors for balancing reserve of generator $i$. Note that $\alpha_{i,u}$ denotes the participation factor  of generator $i$ in response to the RES forecast error at node $ u \in \set{U}$.  Chance constraints \cref{base:upper_gen_cc,base:lower_gen_cc} ensure the power output of conventional generator $i$ under uncertainty does not exceed the upper  or lower   limits $\overline{p}_{G,i}$ and $\underline{p}_{G,i}$  with a probability of $1-\epsilon_g$. Similarly, \cref{base:flowconstraints} ensures that DC power flows computed using function $F(p_{G,i}(\om), p_{U,i}(\om), p_{D,i})$, which maps net nodal injections in power flows, are contained in a convex set of feasible power flows given by $\set{F}$ with a probability of $1-\epsilon_f$. 
Finally,  \cref{base:enerbal} is the nodal power balance constraint given the nodal  demand and  power flow injections  $p_{D,i}$ and  $p_i(F)$.
{\color{black} 
Eq.~\cref{base:balbal} ensures that the procured balancing reserve is sufficient to mitigate $\om$.  
}
Greek letters in parentheses denote dual multipliers.

\subsection{Deterministic Reformulation}

Assuming that $c_i(p_{G,i})$ is  quadratic:
\begin{equation}
    c_i(p_{G,i}(\om)) = c_{2i}(p_{G,i}(\om))^2 + c_{1i}p_{G,i}(\om) + c_{0i}, 
\end{equation}
where  $c_{2i}$,  $c_{1i}$,  $c_{0i}$ are cost coefficients, and using $\RM_0 \equiv \Eptn$ and $\om \sim \mathcal N(0, \Sigma)$, where $\Sigma$ is the covariance matrix of $\om$, \cref{mod:base-cc} has a tractable convex (conic) reformulation, \cite{bienstock2014chance}:
\allowdisplaybreaks
\begin{subequations}
\begin{align}
     \min_{\substack{p_{G,i}, \alpha_i \\ s_{p_{G,i}}}} \quad 
        & \sum_{i\in\set{G}} c_i(g_i) + c_{2i} \norm{\alpha_i^{\top}\Sigmart}_2^2 \label{base_ref:objective}\\
    \text{s.t.}\quad
    (\zeta_i): \quad
        & s_{p_{G,i}} \geq \norm{\alpha_i^\top \Sigmart}_2  && \forall i \in \set{G} \label{base_ref:zeta_i}\\
    (\delta_i^+): \quad
        & p_{G,i} + z_{1-\epsilon_g} s_{p_{G,i}} \leq \overline{p}_{G,i} && \forall i \in \set{G}  \label{base_ref:upper_gen_cc}\\
    (\delta_i^i): \quad
        & -p_{G,i} + z_{1-\epsilon_g} s_{p_{G,i}} \leq -\underline{p}_{G,i} && \forall i \in \set{G}  \label{base_ref:lower_gen_cc} \\
    (\theta): \quad    
        & \tilde{F}_{\epsilon_f}(p_{G}, p_{U}, p_{D}, \alpha) \leq 0 \label{base_ref:flowconstraints} \\
    (\lambda_i): \quad
        &  p_{G,i} + p_{U,i} + p_i(\tilde{F}_{\epsilon_f})  = p_{D,i} && \forall i \in \set{N}\label{base_ref:enerbal}, \\
    (\chi_u): \quad
        & \sum_{i\in\set{G}} \alpha_{i,u} = 1 && \forall u \in \set{U} \label{base_ref:balbal},
\end{align}%
\label{mod:base-cc_ref}%
\end{subequations}%
\allowdisplaybreaks[0]%
where  ${z_{1-\epsilon} = \Phi^{-1}(1-\epsilon)}$ is the quantile function of the standard normal distribution and  $s_{p_{G,i}}$ is an auxiliary decision variable modeling the standard deviation of $p_{G,i}(\om)$. 
{\color{black} 
As explained in \cite{mieth2019distribution}, the  reserve provided by each producer can then be computed as $z_{1-\epsilon_g} s_{p_{G,i}}$, where $s_{p_{G,i}}$  depends on participation factors $\alpha_i$.
}
{\color{black}
(Note that this expression holds even for more general distribution assumptions on $\om$, see  \cite{dvorkin2019chancemarket}).} 
Function $\tilde{F}_{\epsilon_f}(\cdot)$ in \cref{base_ref:flowconstraints} maps the decision variables, parameters, statistical characteristics of $\om$ and security threshold $\epsilon_f$ into a vector of power flows with security margins so that \cref{base_ref:flowconstraints} is equivalent to chance constraint \cref{base:flowconstraints}.

\subsection{Equilibrium Formulation}

The optimization problem  in  \cref{mod:base-cc,mod:base-cc_ref} represents a risk-neutral market operator and has been proven to yield  energy and balancing reserve prices $\lambda_i$ and $\chi_u$, which solve the following equilibrium, \cite{kuang2018pricing,dvorkin2019chancemarket,mieth2019risk}:
\begin{subequations}
\begin{align}
&\begin{Bmatrix*}[l]
    \max_{\substack{p_{G,i}, \alpha_i \\ s_{p_{G,i}}}} \ 
        & \lambda_i p_{G,i} + \chi^{\top} \alpha_i - \Eptn[c_i(p_{G,i}(\om))] \\
        \hfill \text{s.t.} &  \text{\cref{base_ref:zeta_i,base_ref:upper_gen_cc,base_ref:lower_gen_cc}} \\
\end{Bmatrix*},\ \forall i \in \set{G} \label{base_eq:prod_problems} \\
    & \qquad \qquad \qquad \ 
    \text{\color{black} \cref{base_ref:enerbal,base_ref:balbal,base_ref:flowconstraints}} 
    \label{base_eq:market_clearing}
\end{align}%
\label{mod:base_eq}%
\end{subequations}%
where \cref{base_eq:prod_problems} is a profit maximization solved by each  conventional generator (producer)  and  
{\color{black} \cref{base_eq:market_clearing} are the  market-clearing conditions.} 
As shown in \cite{dvorkin2019chancemarket,mieth2019risk,kuang2018pricing}, $\lambda_i$ and $\chi_u$ can be interpreted as  equilibrium energy and reserve prices.

\section[Risk-Averse Chance-Constrained Electricity Market]%
    {Risk-Averse Chance-Constrained\\ Electricity Market}

The optimization in \cref{base_eq:prod_problems} solved by each producer is risk neutral because it assumes average (expected) outcomes of random  $\om$. In practice, however, producers are likely to hedge against the risk of uncertain costs based on their risk perception.
This section considers risk-averse profit maximizing producers endowed with  a \textit{risk measure} $\RM_i$.

\subsection{Coherent Measures of Risk}

Intuitively, a risk measure evaluates an uncertain outcome $\mathbf{Z}$ in terms of an equivalent deterministic outcome $\RM[\mathbf{Z}]$ so that a producer endowed with risk measure $\RM$ is indifferent between accepting uncertain $\mathbf{Z}$ or its certainty equivalent $\RM[\mathbf{Z}]$.
Additionally, a risk measure is called \textit{coherent} if it satisfies some fundamental mathematical properties such as monotonicity, positive homogeneity, translation invariance and convexity, \cite{rockafellar2007coherent}.
For example, the expectation operator $\Eptn$ is a coherent measure of risk, \cite{rockafellar2007coherent}, but neglects the volatility of outcomes, and is therefore associated with risk-neutrality.


Any coherent risk measure can be expressed as, \cite{rockafellar2007coherent}:
\begin{align}
    \RM[\mathbf{Z}] = \sup_{\Prb\in\set{D}} \Eptn_{\Prb}[\mathbf{Z}] \label{eq:risk_measure_reform}
\end{align}
where $\set{D}$ denotes the \textit{risk set} (risk envelope) of $\RM$, i.e. a compact convex set of probability measures, and $\Eptn_{\Prb}$ is the expectation over the probability measure $\Prb$.
Risk set $\set{D}$ uniquely defines $\RM$ and can be structured such that $\sup_{\Prb\in\set{D}} \Eptn_{\Prb}[\mathbf{Z}]$ is identical to specific risk measures, e.g. CVaR, \cite{rockafellar2007coherent}.

\begin{remark}
\label{rem:dis_robust_connection}
Defining a risk measure in terms of a worst-case probability distribution as in \cref{eq:risk_measure_reform} is structurally identical to \textit{distributionally robust optimization} that can be applied to chance constraints \cref{base:upper_gen_cc,base:lower_gen_cc,base:flowconstraints}, see e.g. \cite{dvorkin2019chancemarket}.
This work, however, focuses on the evaluation of the objective, 
{\color{black} i.e. the reformulation of  constraints in \cref{base_ref:upper_gen_cc,base_ref:lower_gen_cc,base_ref:flowconstraints} remains unchanged.
}
\end{remark}

\subsection{Risk-Averse Profit Maximization}
\label{ssec:risk_avers_profit_maximization}

To derive a risk-averse modification of \cref{mod:base-cc_ref}, we define a risk set using a moment ambiguity set, which generally yields tractable convex optimization problems, \cite{delage2010distributionally}.
Thus, the risk set of each producer $i$ is:
\begin{align}
    \set{D}_i = \{\Prb(\om) \in \set{P} \mid \Eptn_{\Prb}[\om] = 0, \Var_{\Prb}[\om] \in \set{S}_i\},
\label{eq:Di_discrete}
\end{align}
where $\set{P}$ is the set of probability distributions  and $\set{S}_i = \{\Sigma_1,...,\Sigma_K\}$ is the set  of  $K$ covariance matrices ($\Sigma_1,...,\Sigma_K$), where $K$  is the same for all producers. 
{\color{black} 
Set $\set{S}_i$, and thus set $\set{D}_i$, captures the belief of producers on the accuracy of RES forecast data and forecasting methods.
Notably, set $\set{D}_i$ is a set of continuous distributions as opposed to discrete polyhedral probability measures in \cite{philpott2016equilibrium,gerard2018risk}, which rely on a given set of pre-described scenarios. 
Hence, using \cref{eq:Di_discrete,eq:risk_measure_reform} yields:}
\begin{equation}
\begin{split}
    &\min_{p_{G,i}, \alpha_i} \quad \sup_{\Prb\in\set{D}_i}  \Eptn_{\Prb}[ c_i(p_{G,i}(\om))] \\
    &\quad = \min_{p_{G,i}, \alpha_i} \quad c_i(p_{G,i}) + \sup_{k=1,...,K} c_{2i} \norm{\alpha_i^{\top} \Sigmartk}_2^2.
\end{split}
\label{eq:wc_cost1}
\end{equation}
Although $\set{D}_i$ as defined in \cref{eq:Di_discrete} is non-convex, solving \cref{eq:wc_cost1} is equivalent to solving the following problem with  convex polyhedral set $\tilde{\set{S}}_i = \conv(\set{S}_i)$, \cite[Section 6.4.2]{boyd2004convex}:
\begin{equation}
    \min_{p_{G,i}, \alpha_i} \quad c_i(p_{G,i}) + \sup_{\Sigma_k \in \tilde{\set{S}}_i} c_{2i} \norm{\alpha_i^{\top} \Sigmartk}_2^2
\label{eq:risk_adjusted_cost_ref}
\end{equation}
and we can define:
\begin{align}
    \tilde{\set{D}}_i = \{\Prb(\om) \in \set{P} \mid \Eptn_{\Prb}[\om] = 0, \Var_{\Prb}[\om] \in \tilde{\set{S}}_i\}
\label{eq:Di_contin}
\end{align}
as the convex counterpart of $\set{D}_i$, which yields the following coherent risk measure:
\begin{equation}
    \RM_i \left[ c_i(p_{G,i}(\om))\right] = \sup_{\Prb\in\tilde{\set{D}}_i}  \Eptn_{\Prb}[ c_i(p_{G,i}(\om))].
\end{equation}

{\color{black} 
Using the epigraph form of \cref{eq:risk_adjusted_cost_ref}, the cost minimization in \cref{mod:base-cc_ref} can be recast as the following risk-averse modification:
\begin{subequations}
\begin{align}
    \min_{\substack{p_{G,i}, \alpha_i \\ s_{p_{G,i}}, t_i}} \quad
        & \sum_{i\in\set{G}} (c_i(p_{G,i}) + t_i) \label{ra_ccopf:objectve}\\
    \text{s.t.} \quad
        & \text{\cref{base_ref:zeta_i,base_ref:upper_gen_cc,base_ref:lower_gen_cc,base_ref:flowconstraints,base_ref:enerbal,base_ref:balbal}} \\
    (\eta_{i,k}): \quad 
        & t_i \geq c_{2i} \norm{\alpha_i^{\top} \Sigmartk}_2^2 \quad \forall \Sigma_k \in \set{S}_i,\, \forall i. \label{ra_ccopf:wc_cost}
\end{align}%
\label{mod:ra_ccopf}%
\end{subequations}%
Similarly, the risk-averse modification of  \cref{base_eq:prod_problems} follows as:
\begin{subequations}
\begin{align}
    \min_{\substack{p_{G,i}, \alpha_i \\ s_{p_{G,i}}, t_i}} \quad
        & \lambda_i p_{G,i} + \chi^{\top} \alpha_i - c_i(p_{G,i}) - t_i \\
    \text{s.t.} \quad                        
        & \text{\cref{base_ref:zeta_i,base_ref:upper_gen_cc,base_ref:lower_gen_cc,ra_ccopf:wc_cost}}.
\end{align}%
\label{mod:ra_proftimax}%
\end{subequations}%
}%
{\color{black} 
Note that  $\Sigma$ in \cref{base_ref:zeta_i} remains unchanged and common for all producers as indicated by Remark~\ref{rem:dis_robust_connection}.
}

\begin{remark}
Unlike in \cref{base_eq:prod_problems}, the risk-averse profit maximization in \cref{mod:ra_proftimax} allows different producers to have different perceptions of the system uncertainty, which can be modeled as different risk attitudes drawn from producer-specific set $\set{D}_i$.
\end{remark}

\section[Risk Trading in the Chance-Constrained Electricity Market]%
{Risk Trading in the Chance-Constrained \\ Electricity Market}
\label{sec:risk_trading}

If producer $i$ is endowed with coherent risk measure $\mathbb{F}_i$ given by risk set $\set{D}_i$ and seeks to maximizes its risk adjusted profit as in \cref{mod:ra_proftimax}, 
{\color{black} 
its decision will be more conservative in the absence of risk-trading opportunities. 
Thus, a risk-incomplete market as in \cref{mod:ra_ccopf} will be less efficient and suffer welfare losses.
} 
This section describes an approach to complete the chance-constrained market with respect to risk by introducing ADS trading.

\subsection{Continuous Risk Trading}
\label{ssec:continous_risk_trading}

ADS as introduced in \cite{arrow1973role} is a common security contract that depends on the outcome of an uncertain asset, which in the case of the chance-constrained electricity market in \cref{mod:ra_ccopf} is the RES forecast error given by  $\om$. Thus, a buyer of the contract pays price $\mu(\om)$ to receive a payment of $1$ for a pre-defined realization of $\om$. Hence, if producer $i$ seeks to receive a payment of $a_i(\om)$ for all possible $\om$, it pays in advance: 
\begin{equation}
    \pi_{a_i} = \int_\Omega \mu(\om) a_i(\om) d\om
\label{eq:ADS_cost}
\end{equation} 
where $\Omega$ is the space of all possible outcomes of random  $\om$. If $a_i(\om) \leq 0$, then producer $i$ sells ADS (i.e. provides  security to the system) and receives the payment of $\pi_{a_i}\leq0$. Otherwise, if  $a_i(\om) \geq 0$, producer $i$ purchases ADS and pays $\pi_{a_i}\geq0$.  
Further, the market must ensure revenue adequacy, i.e. that the amount of ADS purchased and sold match:
\begin{equation}
  (\mu(\om)): \quad  \sum_{i\in\set{G}} a_i(\om) = 0 \quad \forall \om \in \Omega.
\label{eq:ads_market_clearing}
\end{equation}
Given the risk trading model in  \cref{eq:ADS_cost,eq:ads_market_clearing}, each profit-maximizing producer can be modeled as follows:
\allowdisplaybreaks
\begin{subequations}
\begin{align}
    \max_{\substack{p_{G,i}, \alpha_i, a_i(\om) \\ s_{p_{G,i}}, t_i}} \quad
        & \lambda_i p_{G,i} + \chi^{\top} \alpha_i - t_i - \pi_{a_i} \\
    \text{s.t.} \quad                       
        & \text{\cref{base_ref:zeta_i,base_ref:upper_gen_cc,base_ref:lower_gen_cc}} \\
    (\eta_{i,k}): \quad
        & \hspace{-3pt}t_i\!\geq\!\Eptn_{\Prb_k}[c_i(p_{G,i}(\om))]\!-\!\Eptn_{\Prb_k}[a_i(\om)], \ \forall \Prb_k\!\in\!\set{D}_i, \label{ra_proftimax_ads:exp_cost}
\end{align}%
\label{mod:ra_proftimax_ads}%
\end{subequations}%
\allowdisplaybreaks[0]%
where $\pi_{a_i}$ reflects the additional cost or revenue  due to risk trading, as given in \cref{eq:ADS_cost}, and  $\Eptn_{\Prb_k}[a_i(\om)]$ in \cref{ra_proftimax_ads:exp_cost} is the expected ADS cost or revenue  over  probability measure $\Prb_k$. Given \cref{eq:ads_market_clearing} and \cref{mod:ra_proftimax_ads}, extending the risk-averse market-clearing in \cref{mod:ra_ccopf} with risk trading yields:
\allowdisplaybreaks
\begin{subequations}
\begin{align}
    \min_{\substack{p_{G,i}, \alpha_i, a_i(\om) \\ s_{p_{G,i}}, t_i}} \quad
        & \sum_{i\in\set{G}} (c_i(p_{G,i}) + t_i) \label{rt_ccopf:objectve}\\
    \text{s.t.} \quad
        & \text{\cref{base_ref:zeta_i,base_ref:upper_gen_cc,base_ref:lower_gen_cc,base_ref:flowconstraints,base_ref:enerbal,base_ref:balbal,eq:ads_market_clearing,ra_proftimax_ads:exp_cost}}, 
\end{align}%
\label{mod:rt_ccopf}%
\end{subequations}%
\allowdisplaybreaks[0]%
where \cref{eq:ads_market_clearing} enforces the market-clearing condition yielding dual multiplier $\mu(\om)$. Using \cref{mod:rt_ccopf} and under the assumption that set $\set{F}$ is sufficiently large to accommodate injections $p_{G}(\om)$, $p_{U}(\om)$, $p_{D}$ without causing network congestion%
\footnote{This assumption simplifies derivations%
\textcolor{black}{, but the result holds for the congested case if  transmission assets and services  are priced \cite{o2008towards}}.} 
(i.e. energy prices are uniform  $\lambda = \lambda_i$), we prove: 
\begin{prop}
\label{prop:equivalence_to_mu}
    Let $\lambda$, $\chi$, and $\mu(\om)$ be equilibrium energy, balancing, and risk prices, respectively, so that $\{\lambda; \chi_u; \mu(\om); p_{G,i}, \forall i \in \set{G}; \alpha_i, \forall i \in \set{G}; a_i(\om), \forall i \in \set{G}\}$ solves \cref{mod:rt_ccopf}. 
    Then $\mu(\om)$ can be interpreted as a probability measure that solves a risk-neutral equivalent of the risk-averse profit maximization with ADS trading.
\end{prop}

\myproofstart
The market-clearing problem in \cref{mod:rt_ccopf} 
{\color{black}%
remains
}%
convex as long as $a_i(\om)$ is convex in $\om$.  Therefore, KKT conditions can be invoked. 
The Lagrangian function of the profit maximization of each producer in \cref{mod:ra_proftimax_ads}  can be written as:
\allowdisplaybreaks
\begin{align}
    \mathcal{L}_i &= \lambda p_{G,i} + \chi^{\top} \alpha_i - t_i - \pi_{a_i} - \zeta_i(\norm{\alpha_i^\top \Sigmart}_2 - s_{p_{G,i}}) \nonumber \\
                & \quad - \delta_i^+\!(p_{G,i}\!+\!z_{\epsilon} s_{p_{G,i}}\!\!\!-\! \overline{p}_{G,i})   
                \!-\! \delta_i^-\!(-p_{G,i}\!+\!z_{\epsilon} s_{p_{G,i}} \!\!\!+\! \underline{p}_{G,i})   \nonumber  \\
                & \quad - \sum_{k=1}^K \eta_{i,k}(\Eptn_{\Prb_k}[c_i(p_{G,i}(\om)) - a_i(\om)] - t_i) \label{eq:lagrangian_ind}
\end{align}%
\allowdisplaybreaks[0]%
Hence, the resulting optimality conditions for $t_i$, $a_i(\om)$ are:
\allowdisplaybreaks
\begin{align}
\begin{split}
    \frac{\partial\mathcal{L}_i}{\partial t_i} &= -1 + \sum_{k=1}^K \eta_{i,k} = 0
        \quad \Rightarrow \quad  \sum_{k=1}^K \eta_{i,k} = 1  
\end{split}\label{eq:Li_over_ti}\\
\begin{split}
    \frac{\partial\mathcal{L}_i}{\partial a_i(\om)} &= -\mu(\om) + \sum_{k=1}^K \eta_{i,k} f(\om,\sigma_k) = 0 \\
    & \Rightarrow \quad \mu(\om) = \sum_{k=1}^K \eta_{i,k} f(\om,\Sigma_k),
\end{split}\label{eq:Li_over_aom}%
\end{align}%
\allowdisplaybreaks[0]%
where  $f(\om,\Sigma_k)$ denotes  the probability density function  of a multivariate, zero-mean distribution with covariance $\Sigma_k$.
{\color{black} 
Note that for the derivation of \cref{eq:Li_over_aom} we used:
\allowdisplaybreaks
\begin{align}
    \frac{\partial \pi_{a_i}}{a_i(\om)} 
        &= \frac{\partial}{\partial a_i(\om)} \int_\Omega \mu(\om) a_i(\om) d\om  = \mu(\om),\\
    \frac{\partial}{\partial a_i(\om)}\!\Eptn_{\Prb_k}[a_i(\om)] 
        \!&=\!\frac{\partial}{a_i(\om)}\!\int_\Omega\!\!a_i(\om) f(\om,\Sigma_k)d\om =\! f(\om,\Sigma_k).
\end{align}%
\allowdisplaybreaks[0]%
}%
Conditions \cref{eq:Li_over_ti,eq:Li_over_aom} lead to two relevant observations:
\begin{enumerate}[(O1), topsep=0.5ex, itemsep=0.5ex]
    \item Dual multiplier $\mu(\om)$ in \cref{eq:ads_market_clearing} is a probability measure as it is the weighted average of $K$ probability density functions with zero means and covariance matrices $\Sigma_1,...,\Sigma_k$. In other words,  random  $\mathbf{Z}(\om)\sim\mu(\om)$ has the expected value of $\Eptn_{\mu}[\mathbf{Z}(\om)] = 0$ and the variance of $\Var_\mu[\mathbf{Z}(\om)] = \sum_{k=1}^K \eta_{i,k} \Sigma_k$.
    
    \item Since $\tilde{S}_i$ is a convex set, condition \cref{eq:Li_over_ti} ensures that ${\sum_{i=1}^K \eta_{i,k} \Sigma_k \in \tilde{S}_i}$ and thus $\mu(\om) \in \tilde{\set{D}}_i$.
\end{enumerate}
The set of optimal decisions $\{\lambda; \chi_u; \mu(\om); p_{G,i}, \forall i \in \set{G}; \alpha_i, \forall i \in \set{G}; a_i(\om), \forall i \in \set{G}\}$  maximize $\mathcal{L}_i$  given in \cref{eq:lagrangian_ind}.  Using observation O1,  the fifth term in \cref{eq:lagrangian_ind} recasts  as:
\allowdisplaybreaks
\begin{align}
    & \sum_{k=1}^K \eta_{i,k}(\Eptn_{\Prb_k}[c_i(p_{G,i}(\om)) - a_i(\om)]) \nonumber \\
    & \quad = \sum_{k=1}^K \eta_{i,k} \int_{\Omega}[c_i(p_{G,i}(\om)) - a_i(\om)]f(\om,\Sigma_k)d\om \nonumber  \\
    & \quad = \int_{\Omega}[c_i(p_{G,i}(\om)) - a_i(\om)]\sum_{k=1}^K\eta_{i,k} f(\om,\Sigma_k)d\om \label{eq:risk_neutral_equiv} \\
    & \quad = \int_{\Omega}[c_i(p_{G,i}(\om)) - a_i(\om)]\mu(\om)d\om \nonumber  \\
    & \quad = \Eptn_{\mu}[c_i(p_{G,i}(\om))] - \pi_{a_i}. \nonumber 
\end{align}%
\allowdisplaybreaks[0]%
Substituting  \cref{eq:risk_neutral_equiv} in \cref{eq:lagrangian_ind}  leads to:
\begin{equation}
   \mathcal{L}_i =  p_{G,i} + \chi \alpha_i - \Eptn_{\mu}[c_i(p_{G,i}(\om))] - y_i^{\delta} - y_i^{\zeta},
\label{eq:risk_neutrak_profit_equiv}
\end{equation}
where $y_i^{\delta}$, $y_i^{\zeta}$ denote the terms related to duals $\delta_i$, $\zeta_i$ in \cref{eq:lagrangian_ind}.  Hence, \cref{eq:risk_neutrak_profit_equiv} is a risk-neutral equivalent, evaluated with respect to probability measure $\mu(\om)$, of the risk-averse profit of producer~$i$ participating in risk trading with ADS.
\myproofend

Given Proposition~\ref{prop:equivalence_to_mu}, the optimization of individual producers in \cref{mod:ra_proftimax_ads} is related to the risk-averse chance-constrained electricity market with ADS trading in \cref{mod:rt_ccopf}:
\begin{prop}
\label{prop:wc_socialplaner}
    Let $\lambda$, $\chi_u$, and $\mu(\om)$ be equilibrium energy, balancing, and risk prices so that $\{\lambda; \chi_u; \mu(\om); p_{G,i}, \forall i \in \set{G}; \alpha_i, \forall i \in \set{G}; a_i(\om), \forall i \in \set{G}\}$ solves problem \cref{mod:rt_ccopf}. 
    Assuming that risk sets $\tilde{\set{D}}_i, i \in \set{G}$ are non-disjoint, i.e. $\bigcap_{i \in \set{G}}\tilde{\set{D}}_i \neq \emptyset$, then these prices and allocations solve the risk-averse chance-constrained market with risk trading with $\tilde{\set{D}}_0 = \bigcap_{i \in \set{G}}\tilde{\set{D}}_i$ and worst case probability measure $\mu(\om)$.
\end{prop}
\myproofstart
Given the optimal solution for each producer, it follows from the complementary slackness of \cref{ra_proftimax_ads:exp_cost}:
\begin{equation}
    \eta_{i,k}(\Eptn_{\Prb_k}[c_i(p_{G,i}(\om)) - a_i(\om)] - t_i) = 0.
    \label{eq:exp_cost_complementary_slackness}
\end{equation}%
{\color{black} 
By summing \cref{eq:exp_cost_complementary_slackness} over all $i$ and $k$, comparing with \cref{eq:risk_neutral_equiv}, and using \cref{eq:ads_market_clearing} to eliminate $\pi_{a_i}$, we write:
}
\allowdisplaybreaks
\begin{align}
    \sum_{i\in\set{G}} t_i 
        & = \sum_{i\in\set{G}} \sum_{k=1}^K \eta_{i,k}(\Eptn_{\Prb_k}\left[c_i(p_{G,i}(\om)) - a_i(\om)\right]  \nonumber \\
        & = \Eptn_{\mu}\left[\sum_{i\in\set{G}}c_i(p_{G,i}(\om)\right]. \label{eq:sum_over_ti}
\end{align}%
\allowdisplaybreaks[0]%
Also, since \cref{ra_proftimax_ads:exp_cost} is a convex epigraph, it follows:
\begin{equation}
\begin{split}
    t_i &= \max_{\Prb \in \set{D}_i}     \Eptn_{\Prb_k}[c_i(p_{G,i}(\om)) - a_i(\om)] \\
        &= \max_{\Prb \in \tilde{\set{D}}_i}     \Eptn_{\Prb_k}[c_i(p_{G,i}(\om)) - a_i(\om)].
\label{eq:t_i}
\end{split}%
\end{equation}%
Given \cref{eq:t_i}, term  $ \sum_{i\in\set{G}} t_i $ in \cref{eq:sum_over_ti}  can also be written as:
\allowdisplaybreaks
\begin{align}
    \sum_{i\in\set{G}} t_i & =  \sum_{i\in\set{G}} \max_{\Prb_k \in \tilde{\set{D}}_i} \Eptn_{\Prb_k}[c_i(p_{G,i}(\om)) - a_i(\om)] \nonumber \\ 
    & \stackrel{\text{\circled{\footnotesize A}}}{\geq} \max_{\Prb \in \bigcap_{i \in \set{G}} \tilde{\set{D}}_i} \Eptn_{\Prb}\left[\sum_{i\in\set{G}}c_i(p_{G,i}(\om)) - a_i(\om)\right] \label{eq:expectation_interception} \\
    & \stackrel{\text{\circled{\footnotesize B}}}{=} \max_{\Prb\in \bigcap_{i \in \set{G}} \tilde{\set{D}}_i}  \Eptn_{\Prb}\left[\sum_{i\in\set{G}}c_i(p_{G,i}(\om))\right],\nonumber
\end{align}%
\allowdisplaybreaks[0]%
where transition $\circled{\footnotesize A}$ is due to the replacement of individual risk sets $\set{D}_i$ with the intersection of all risk sets $\tilde{\set{D}}_0 = \bigcap_{i \in \set{G}}\tilde{\set{D}}_i$ and transition $\circled{\footnotesize B}$ is due to the market-clearing ADS condition in \cref{eq:ads_market_clearing}. 
Since $\mu(\om) \in \tilde{\set{D}}_i, \forall i\in \set{G}$ and $\tilde{\set{D}}_0 \neq \emptyset$, due to observation O2 above, \cref{eq:sum_over_ti,eq:expectation_interception} yield
\begin{equation}
     \Eptn_{\mu}\!\left[\sum_{i\in\set{G}}c_i(p_{G,i}(\om))\right]\!=\!\max_{\Prb_k \in \tilde{\set{D}}_0} \! \Eptn_{\Prb_k}\!\left[\sum_{i\in\set{G}}c_i(p_{G,i}(\om))\right],
\end{equation}
showing that $\mu(\om)$ is the worst-case probability measure for the risk-averse market with risk trading. 
\myproofend

\begin{remark}
Propositions \ref{prop:equivalence_to_mu} and \ref{prop:wc_socialplaner} hold if  \cref{mod:rt_ccopf} has binding constraints in \cref{base_ref:flowconstraints}, which can be proven analogously.  
\end{remark}

\subsection{Discrete Risk Trading}
\label{ssec:discrete_risk_trading}

{\color{black} 
Recall that Section~\ref{ssec:continous_risk_trading} defines ADS as continuous over $\om$, which leads to an infinite-dimensional problem in \cref{{mod:rt_ccopf}} and obstructs tractable computations and designing practical risk contracts. To overcome these caveats, the probability space of $\om$ can be discretized to consider contracts for discrete events.
}
Hence,   consider the system-wide (aggregated) RES forecast error  given as $\mathbf{O} = e^{\top}\om$ with mean $\Eptn_{\Prb_k}[\mathbf{O}] = 0$ and variance ${\Var_{\Prb_k}[\mathbf{O}] =  e^{\top}\Sigma_k e \eqqcolon \sigma_k^2}$, where $e$ is the vector of ones of appropriate dimensions. The probability space of $\mathbf{O}$ can then be divided into $W$ events $w=1,...,W$, where each event is a closed interval given by   $\set{W}_w = [l_w, u_w]$ so that $\bigcup_{w=1}^W \set{W}_w  = \mathbb{R}$.
These intervals are sequential such that $l_1 = -\infty$, $u_W = \infty$ and $u_w = l_{w+1}, w=1,...,W\!-\!1$. Using this discretization, the probability of each discrete outcome is defined by $\Prb_k$ as:
\allowdisplaybreaks
\begin{align}
    P_w(\sigma_k) \coloneqq
    \Prb_k[\mathbf{O} \in \set{W}_w] 
        &= \Prb_k[(\mathbf{O} \leq u_w)\cap (\mathbf{O} \geq l_w) ]  \nonumber\\
        & = \int_{l_w}^{u_w} \!\! f(x,\sigma_k)dx 
\label{eq:event_probabilites}
\end{align}%
\allowdisplaybreaks[0]%
and can be pre-computed for all $w=1,...,W$ and $k=1,...,K$. Using the discrete space notation,  \cref{eq:ADS_cost} recasts as:
\begin{align}
    \pi_{a_i} = \sum_{w=1}^W \mu_w a_{i,w},
\label{eq:p_ai_discrete}
\end{align}
where $a_{i,w} \in \mathbb{R}$. Next, using  \cref{eq:event_probabilites},  the expected cost or payment $a_i(\om)$ under $\Prb_k$ can be computed as:
\begin{equation}
\begin{split}
    \Eptn_{\Prb_k}[a_i(\om)] 
        & = \sum_{w=1}^W a_{i,w} P_w(\sigma_k).
\end{split}%
\label{eq:ai_expectation_discrete}%
\end{equation}%
Finally, using  \cref{eq:p_ai_discrete,eq:ai_expectation_discrete} and
{\color{black} the discrete-space equivalent of \cref{eq:Li_over_aom}, i.e. the optimality condition for $a_{i,w}$,
}%
the discrete-space equivalent of $\mu(\om)$ is computed as:
\begin{equation}
\begin{split}
  \mu_{w} 
        & = \sum_{k=1}^K \eta_k P_w(\sigma_k) 
         = \sum_{k=1}^K \eta_{i,k} \int_{l_w}^{u_w} f(x,\sigma_{i,k})dx \\
        & = \int_{l_w}^{u_w} \sum_{k=1}^K \eta_{i,k} f(x,\sigma_{i,k})dx \quad \forall i\in\set{G},
\end{split}
\label{eq:mu_discrete}
\end{equation}
where $\sigma_{i,k} = \norm{e^{\top}\Sigmartk}_2$ with $\Sigma_k \in \set{S}_i$.
Hence, due to \cref{eq:mu_discrete},  $\mu_{w}$ retains the interpretation of $\mu(\om)$ from observation O1 of Proposition~\ref{prop:equivalence_to_mu}. Indeed,  a random variable with probability density function $\sum_{k=1}^K \eta_{i,k} f(x,\sigma_{i,k})$ has variance
$\norm{e^{\top}(\sum_{k=1}^K\eta_{i,k}\Sigma_k)^{\nicefrac{1}{2}}}_2^2=e^{\top}(\sum_{k=1}^K\eta_k\Sigma_k)e$, it follows that $\Var_\mu(\mathbf{O})=e^{\top}(\sum_{k=1}^K\eta_{i,k}\Sigma_k)e.$ Using this result and \cref{eq:p_ai_discrete,eq:ai_expectation_discrete,eq:mu_discrete,eq:event_probabilites}, a discrete modification of the risk-averse chance-constrained electricity market  with risk trading in \cref{mod:rt_ccopf} is:
\allowdisplaybreaks
\begin{subequations}
\begin{align}
    \min_{\substack{p_{G,i}, \alpha_i, a_{i,w} \\ s_{p_{G,i}}, t_i}} \quad
        & \sum_{i\in\set{G}} (c_i(p_{G,i}) + t_i) \label{rt_ccopf_discrete:objectve}\\
    \text{s.t.} \quad
        & \text{\cref{base_ref:zeta_i,base_ref:upper_gen_cc,base_ref:lower_gen_cc,base_ref:flowconstraints,base_ref:enerbal,base_ref:balbal}} \\
    (\eta_{i,k}): \quad 
        & t_i \!\geq  c_{2i} \norm{\alpha_i^{\top} \Sigmartk}_2^2 \!+\!\! \sum_{w=1}^W a_{i,w} \!P_w\left(\sigma_{i,k}\right), \nonumber \\ 
        & \hspace{3.5cm} \forall \Sigma_k \in \set{S}_i ,\, \forall i \label{rt_ccopf:wc_cost} \\
    (\mu_w): \quad
        & \sum_{i\in\set{G}} a_{i,w} = 0, \quad \forall w=1,...,W.
    \label{rt_ccopf:ads_market_clearing}
\end{align}%
\label{mod:rt_ccopf_dicrete}%
\end{subequations}%
\allowdisplaybreaks[0]%
Since the discrete representation of ADS contracts in \cref{mod:rt_ccopf_dicrete} is a special case of the infinite-dimensional representation in \cref{mod:rt_ccopf}, the results of Propositions \ref{prop:equivalence_to_mu} and \ref{prop:wc_socialplaner}  hold for \cref{mod:rt_ccopf_dicrete}. 

\subsection{Price Analysis with Risk Trading}
\label{sec:price_analysis}

Using the risk-averse chance-constrained electricity market  with discrete risk trading in \cref{mod:rt_ccopf_dicrete}, this section analyzes resulting energy, balancing reserve and risk prices as follows:
\begin{prop}
\label{prop:energy_prices}
Consider the risk-averse chance-constrained market with risk trading in \cref{mod:rt_ccopf_dicrete}. Let $\lambda_i$, $\chi_u$ and $\mu_w$ be the dual multipliers of the active power balance \cref{base_ref:enerbal}, the reserve sufficiency constraint \cref{base_ref:balbal} and the ADS market-clearing constraint \cref{rt_ccopf:ads_market_clearing}. Then $\mu_w$ is given by \cref{eq:mu_discrete} and  $\lambda_i$, $\chi_u$  are:
\allowdisplaybreaks
\begin{align}
    \hspace{-2.5mm} \lambda_i & = 
        2 c_{2i}p_{G,i} + c_{i1} + (\delta_i^+ - \delta_i^-) + y_{p_{G,i}}(\theta) \label{eq:lambda_rt} \\
    \hspace{-2.5mm}  \chi_u & \!=\! \frac{1}{|\set{G}|} 
        \sum_{i\in\set{G}}\!\!\left(
        \!\!2c_{2i}\alpha_i^{\top}[\overline{\Sigma}_i]_u\!+\!z_{1-\epsilon_g} \delta_i \frac{\alpha_i^{\top}[\Sigma]_u}{s_{G,i}}\!+\! y_{\alpha_{i,u}}\!(\theta)\!\!\right)\!, \label{eq:chi_rt}
\end{align}%
\allowdisplaybreaks[0]%
where $y_{p_{G,i}}(\theta) \coloneqq \theta^{\top}\frac{\partial\tilde{F}_{\epsilon_f}}{\partial p_{G,i}}$, $y_{\alpha_{i,u}}(\theta)\coloneqq\theta^{\top}\frac{\partial\tilde{F}_{\epsilon_f}}{\partial \alpha_{i,u}}$, $\overline{\Sigma}_i \coloneqq (\sum_{k=1}^K \eta_{i,k} \Sigma_k \mid \Sigma_k\in\set{S}_i$), $\delta_i \coloneqq \delta_i^+ + \delta_i^-$, $[X]_u$ is the vector of elements in the  $u$-th column of matrix $X$, and $s_{G,i} = \norm{\alpha_i^{\top}\Sigmart}_2$, i.e. the standard deviation of $p_{G,i}(\om)$. 
\end{prop}
\myproofstart
Let $\mathcal{L}$ be the Lagrangian function of \cref{mod:rt_ccopf_dicrete}, its first-order optimality conditions  for $p_{G,i}$, $\alpha_{i,u}$, $s_{p_{G,i}}$ and $a_{i,w}$ are:
\allowdisplaybreaks
\begin{subequations}
\begin{align}
\begin{split}
    \frac{\partial \mathcal{L}}{\partial p_{G,i}}
        & = 2 c_{2i}p_{G,i} + c_{i1} + (\delta_i^+ - \delta_i^-)  \\
        & + y_{p_{G,i}}(\theta) - \lambda_i = 0,\ \forall i \in \set{G}
\end{split}\label{eq:l_over_pgi}\\
\begin{split}
    \frac{\partial \mathcal{L}}{\partial \alpha_{i,u}}    
        &= 2c_{2i}\alpha_i^{\top}[\overline{\Sigma}_i]_u + \zeta_i \frac{\alpha_i^{\top}[\Sigma]_u}{\norm{\alpha_i^{\top}\Sigmart}_2} \\
        & +  y_{\alpha_{i,u}}(\theta) -\chi_u = 0,\ \forall i \in \set{G},\ \forall u \in \set{U}
\end{split}\label{eq:l_over_alphaiu}\\
\begin{split}
    \frac{\partial \mathcal{L}}{\partial s_{p_{G,i}}}
        & = -\zeta_i + \delta_i^+ z_{1-\epsilon_g} + \delta_i^- z_{1-\epsilon_g} = 0,\ \forall i \in \set{G}
\end{split}\label{eq:l_over_si} \\
\begin{split}
    \frac{\partial \mathcal{L}}{\partial a_{i,w}}
        & = \sum_{k=1}^K \eta_{i,k}P_w(\sigma_{k,i}) - \mu_w = 0
\end{split}\label{eq:l_over_aiw}
\end{align}%
\end{subequations}%
\allowdisplaybreaks%
{\color{black}Expressions \cref{eq:lambda_rt} and \cref{eq:chi_rt}} follow immediately from  \cref{eq:l_over_aiw} and \cref{eq:l_over_pgi}, respectively.
Expressing $\zeta_i$  from \cref{eq:l_over_si} and summing over all $i\in\set{G}$ in \cref{eq:l_over_alphaiu} yields \cref{eq:chi_rt}.
\myproofend

Notably,  energy prices in \cref{eq:lambda_rt} are driven by cost coefficients  of $c_i(\cdot)$ and do not explicitly depend on random $\om$, risk set $\mathcal D_i$ and tolerance to chance constraint violations  $\epsilon_g$. On the other hand, the balancing reserve price in \cref{eq:lambda_rt}  depends on $\om$ (via parameter $\Sigma$),  $\mathcal D_i$ (via parameter $\overline{\Sigma}_i$) and $\epsilon_g$. Finally, risk prices in \cref{eq:mu_discrete} depends on the degree of discretization $W$, which affects interval limits $l_w$ and $u_w$, and individual risk perception given by set $\mathcal D_i$ (via parameter $\sigma_{i,k}$).

\section{Case Study}
\label{sec:case_study}

\begin{figure}
    \centering
    \includegraphics[width=1\linewidth]{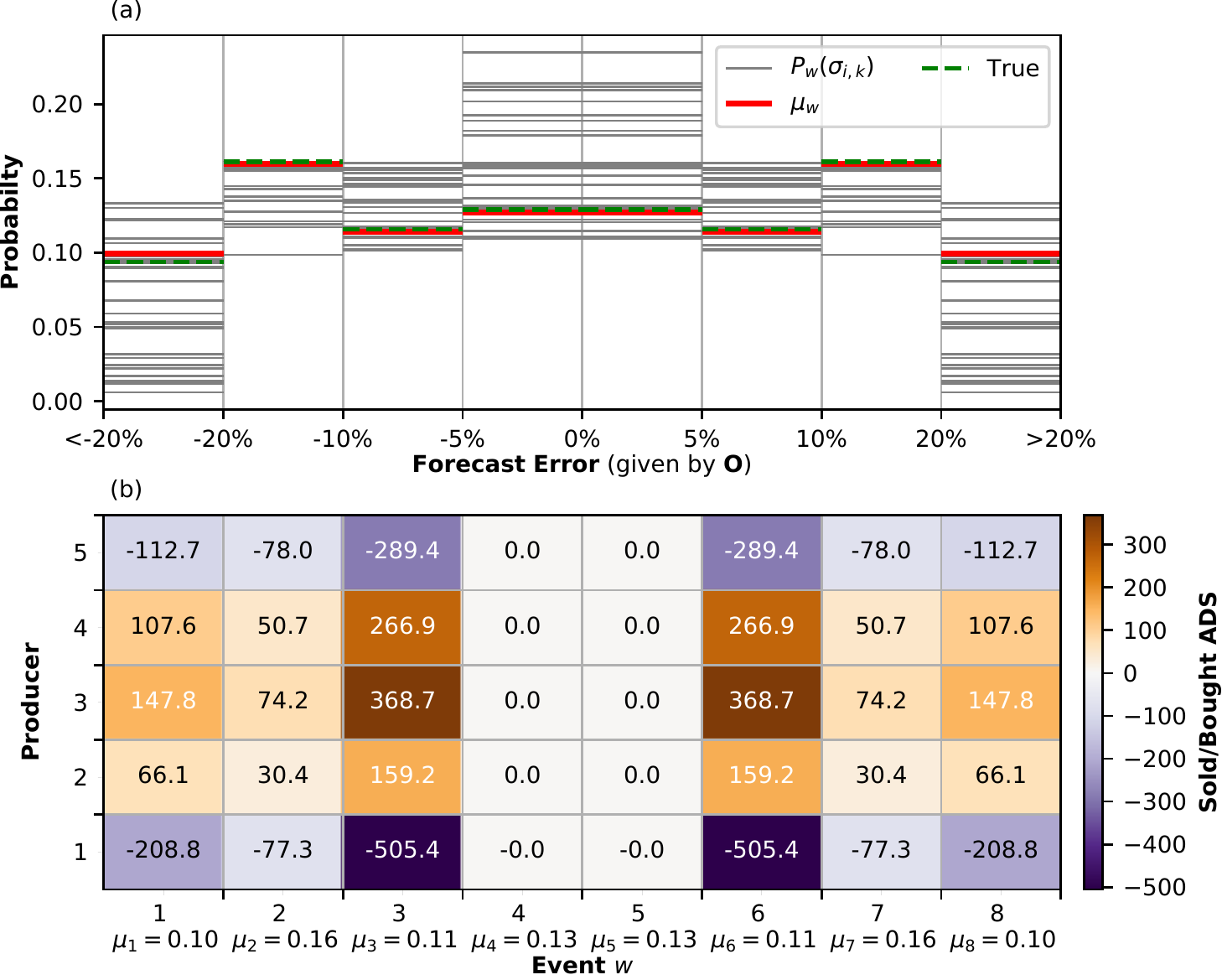}
    \caption{
    {\color{black}
    Risk trading results in the RT case: (a) itemizes the event probabilities $P_w(\sigma_{i,k})$, see \cref{eq:event_probabilites},  drawn from all individual risk sets (shown in thin gray lines) relative to the ``common"   distribution (dashed green line) and the ADS prices $\mu_w$ (solid red line); (b) itemizes the ADS trades, where negative (purple) values indicate a producer selling ADS and positive (orange) values indicate a producer buying ADS. The columns in both (a) and (b) reflect the events with breakpoints indicated on the x-axis of (a) and event numbers as indicated on the x-axis of (b).}
    }
    \label{fig:ads_results}
\end{figure}

{\color{black}
We conduct a case study to illustrate some of the theoretical results of the paper by comparing the \textit{ex ante} outcomes of the risk-averse chance-constrained electricity market without risk trading (``NO-RT''), as formulated in \cref{mod:ra_ccopf}, and with risk trading (``RT''), as formulated in \cref{mod:rt_ccopf_dicrete}.
%
We construct a data set that includes five conventional producers with parameters \unit[$c_{1i} = \{10,7,7,15,17\}$]{\$/MW}, $c_{2i} = 0.1c_{1i}, \forall i \in \set{G}$, \unit[$\overline{p}_{G,i}=\{30,10,10,25,25\}$]{MW} and $\underline{p}_{G,i}=0, \forall i \in \set{G}$, and five undispatchable stochastic RES producers. The total system demand is $\sum_{i\in\set{N}}p_{D,i}=\unit[100]{MW}$ and forecasted RES production is $p_{U,i}=\unit[5]{MW}, \forall i \in \set{U}$.
The risk sets $D_i$ defined by set $S_i$, see \cref{eq:Di_discrete}, of the individual producers are constructed with $K=10$ as follows. 
Each producer $i$ has a set $\set{S}_i$ of $K-1$ covariance matrices that reflect their individual risk perception. 
We randomly generate these sets with the 
standard deviation of $\om_i$ between $0$ to  $0.4 p_{U,i}$ and the correlation between 0 to $0.5$.
Additionally, we assume there exists a ``common'' covariance defined such that all $\om_i$ have a standard deviation of $0.2 p_{U,i}$ and no correlation.  
This common covariance matrix is added to all $\set{S}_i$ and can, for example, reflect information provided by the market operator or some third-party forecast provider.
We create eight ADS events by discretizing the probability space of $\bf{O}$ in eight intervals using  breakpoints $[ -0.2, -0.1, -0.05, 0, 0.05, 0.1, 0.2]$, as explained in Section~\ref{ssec:discrete_risk_trading} and  shown on the x-axis of Fig.~\ref{fig:ads_results}(a). 
The code and data is available in \cite{risk_trading_code}.
}

{\color{black}
For this data set, the RT case reduces the risk-adjusted system cost by 0.2\% relative to the NO-RT case. 
Notably, the energy cost component (\unit[4,656.50]{\$}) and energy prices (\unit[62.09]{\$}) are the same in both cases, but the balancing reserve cost component is reduced by \unit[11]{\%} (from \unit[6.17]{\$} to \unit[5.52]{\$}).  
Similarly, generation levels $p_{G,i}$ remain unchanged for both cases (see Table~\ref{tab:alphas}). On the other hand, the introduction of ADS trading changes the balancing reserve provision ($\alpha_{i,u}$) and its prices ($\chi_u$), as shown in Table~\ref{tab:alphas}, which is influenced by different risk beliefs of market participants.
}

\begin{table}[t]
\setlength{\tabcolsep}{2.5pt}
\caption{Power Outputs and Balancing Participation Factors}
\label{tab:alphas}
    \centering
    \begin{tabular}{c|c|ccccc|ccccc}
    \toprule
     &  & \multicolumn{5}{c|}{ $\alpha_{i,u}$ in the RT case} & \multicolumn{5}{c}{$\alpha_{i,u}$  in the NO-RT case} \\
     & $p_{G,i}$ & $u=1$ & 2 &  3 &  4 &  5 &  1 & 2 &  3 &  4 &  5 \\
    \midrule
        $i=1$ & 26.04 & 0.09 &      0.18 &      0.26 &      0.34 &      0.21 &      0.19 &      0.19 &      0.23 &      0.31 &      0.24 \\
        2 & 10.00 & 0.41 &      0.30 &      0.40 &      0.10 &      0.17 &      0.39 &      0.26 &      0.32 &      0.10 &      0.25 \\
        3 & 10.00 & 0.31 &      0.28 &      0.06 &      0.33 &      0.31 &      0.37 &      0.27 &      0.12 &      0.28 &      0.23 \\
        4 & 15.70 & 0.01 &      0.14 &      0.14 &      0.12 &      0.26 &      0.04 &      0.15 &      0.18 &      0.14 &      0.20 \\
        5 & 13.36& 0.18 &      0.10 &      0.15 &      0.11 &      0.06 &      0.02 &      0.13 &      0.15 &      0.17 &      0.08 \\
    \midrule
    $\chi_u$ & -- &  1.24 & 1.54 & 0.72 & 0.69 & 1.31 & 0.91 & 1.47 & 1.31 & 1.34 & 1.11 \\
    \bottomrule
    \multicolumn{12}{l}{\color{black}
    (Indices $i$ relate to producers, indices $u$ relate to uncertain RES.)}
    \end{tabular}
\end{table}

{\color{black}
Fig.~\ref{fig:ads_results} summarizes the discrete events and resulting risk trades. Each column in Fig.~\ref{fig:ads_results} reflects one event, numbered on the x-axis of Fig.~\ref{fig:ads_results}(b) and with the interval breakpoints shown on the x-axis of Fig~\ref{fig:ads_results}(a).
ADS trading outcomes are itemized in Fig.~\ref{fig:ads_results}(b), where negative and positive values indicate ADS selling and purchasing producers, respectively. 
Due to the symmetry of the RES uncertainty distributions, the ADS trading outcomes are also symmetric. Note that producers 1 and 5 are security providers and producers 2-4 are security takers. Specifically, in the NO-RT case, producer 5 expects to attain a greater profit by providing less balancing reserve to RES $u=1$ than in the RT case. In other words, when producer 5 can hedge its risk via ADS trade, it is incentivized to procure more balancing reserve for RES $u=1$.
The risk-aversion also affects the  ADS prices in Fig.~\ref{fig:ads_results}(b)  given by dual $\mu_w$ of the ADS market-clearing constraint \cref{eq:ads_market_clearing} for each event. As shown in Fig.~\ref{fig:ads_results}(a), the values of risk prices $\mu_w$ in Fig.~\ref{fig:ads_results}(b), match  the ``common'' event probabilities. That is, $\mu_w$ is indeed a probability measure, as in Proposition~\ref{prop:equivalence_to_mu}, and captures the risk perception at the intersection of all risk sets $\tilde{\set{D}}_i$, as in Proposition~\ref{prop:wc_socialplaner}.
}

\section{Conclusion}

This paper has developed a risk-averse modification of the chance-constrained electricity market proposed in   \cite{dvorkin2019chancemarket,mieth2019risk,mieth2019distribution} by completing it with ADS-based risk trading.  By discretizing the outcome space of the system uncertainty, we formulated practical ADS contracts that lead to a computationally tractable market-clearing optimization with risk trading. This optimization reduces the system operating cost relative to the case with no risk trading and  produces energy, balancing reserve and risk prices. In particular, both qualitative and quantitative analyses indicate that system uncertainty and risk parameters do not explicitly affect the energy prices, but explicitly contribute to the formation of the balancing reserve and risk prices.

\bibliographystyle{IEEEtran}
\bibliography{literature}

\clearpage

\end{document}